\documentclass[12pt]{article} 
\usepackage{english}

\begin{document} 

\thispagestyle{empty} 

\begin{flushright}

IFUP-TH 56/96 \\
September 1996                                                        

\end{flushright} 

\bigskip\bigskip                                                               

\centerline{{\sl Reply to A. Patrascioiu's and E. Seiler's 
comment on our paper}}
\begin{center}
{\bf \Large{The two-phase issue in the $O(n)$ non-linear }}
\vskip2mm 
{\bf \Large {$\sigma$-model: a Monte-Carlo study}}
\end{center}
\vskip 1.0truecm
\centerline{\bf
B. All\'es, A. Buonanno and G. Cella}
\vskip5mm
\centerline{\it Dipartimento di Fisica and INFN,}
\vskip 2mm
\centerline{\it Piazza Torricelli 2, 56126 Pisa, Italy}

\vskip 2cm

Our paper \cite{stlouis} has motivated a comment
\cite{comment} by A. Patrascioiu and E. Seiler 
which we reply in this note. The remarks in \cite{comment} 
concern three statements that the authors select from our paper:

{\it i)} {\sl ``The results for $O(8)$ 
support the asymptotic freedom scenario''}.

The authors in \cite{comment} recall that at small 
$\tilde\beta \equiv \beta/N$ the $1/N$ expansion is Borel summable
\cite{frohlich}.
No rigorous proof has been given concerning the behaviour 
of the series at large $\tilde\beta$. 
In \cite{comment} it is not shown
if our working $\beta$'s ($\beta \sim 4.6-6.5$) lie in
the ``small'' or ``large'' $\tilde\beta$ region when $N=8$
but at least we can say that if the results
from our Monte Carlo data fit so nicely the $PT$ predictions for the $O(8)$
model (within few per mille for the
mass gap and susceptibility predictions, see \cite{inprogress}) there is
reasonable room to think that the exact $O(8)$ model (what we simulate)
has a critical point at $g=0$ and the set of $PT$ predictions are correct.
If our data had to agree with some prediction other than the 
$PT$ set of predictions for the $O(8)$ model then
the small difference between our result for the mass gap and the
P. Hasenfratz et al. \cite{hasenfratz} prediction, 0.5\%
(compared for instance 
to the difference of almost 30\% between the $n=8$ and the
$n=\infty$ predictions) would become an intriguing challenge.
Considering it as an accident is a matter of feelings.

\newpage

{\it ii)} {\sl ``Assuming finite-size scaling (FSS), it has been shown 
that $O(3)$ presents
asymptotic scaling starting from $\xi=10^5$''}.

This statement is true: {\bf Assuming} FSS it has been shown that
the $O(3)$ model presents asymptotic scaling within few per cent at
large correlation lengths. We are aware of the 
validity of $PT$ whenever the limit 
$L/\xi \to 0$ holds, where $L$ is the 
lattice size and $\xi$ any correlation length,
(see for instance \cite{beccaria}). For this reason we made our simulations
at large values of the previous ratio, $L/\xi \sim 7-10$. Therefore
the second comment of \cite{comment} does not apply to us.

\vskip 3mm

{\it iii)} {\sl ``The $O(3)$ model with Symanzik action 
does not show KT behavior''}.

Strangely enough we have not written this sentence in our paper
\cite{stlouis}. Maybe the authors had in mind some of the following
sentences that do appear in the paper:
\begin{enumerate}
\item[(1.)] {\sl ``If the constancy of $R_{KT}$ ... is a genuine physical
effect, then also for the Symanzik action we should see such a behaviour''}.
\item[(2.)] {\sl ``... our data [for the ratio $R_{KT}$] are not constant''}.
\item[(3.)] {\sl ``... our data [for $O(3)$] do not support either $KT$ or
$PT$''}.
\end{enumerate}

(1.) The tree-level improved 
Symanzik action (although invented in the $PT$ context) is as good
as any other action for describing the $O(n)$ models on the lattice
and it does not assume the validity of $PT$.
Both the Lipschitz action (see for instance \cite{statphys}) 
and the Symanzik action are in the same universality class
as the standard action.

(2.) Figure 1 in \cite{stlouis} clearly shows that the $KT$
ratio is not constant, in contrast to what happens for the 
standard action \cite{summer85}. We stress the fact that our data
have better statistics. Clearly a similar high-statistics 
simulation for the standard action is worth doing.

(3.) We clearly say in \cite{stlouis} that the data for the mass, 
magnetic susceptibility and ratio $R_{PT}$ for the $O(3)$ model 
satisfy the $PT$ predictions
only within $15\% $, \cite{inprogress}. 
This fact, together with point (2.), led us to 
conclude that our data do not support
either $KT$ or $PT$. We cannot say more than this for the $O(3)$ model.

\end{document}